\documentclass[aip,jcp,preprint,showkeys,superscriptaddress,longbibliography,noeprint]{revtex4-1}
\usepackage[utf8]{inputenc}
\usepackage[english]{babel}
\makeatletter\AtBeginDocument{\let\@elt\relax}\makeatother

\usepackage{systeme}
\usepackage[fleqn]{amsmath}
\usepackage{amssymb}
\usepackage[separate-uncertainty=true, multi-part-units=single]{siunitx}
\usepackage[inline,shortlabels]{enumitem}

\newcommand\fh[1]{{\color{black} #1}}

\newcommand\roya[1]{{\color{black} #1}}

\newcommand{\VEC}[1]{\boldsymbol{#1}}
\newcommand\kB{k_\mathrm{B}}

\renewcommand\phi{\varphi}
\renewcommand\rho{\varrho}

\newcommand{\Res}{\text{Res}}
\newcommand\AdResS{AdResS} 

\usepackage{graphicx,grffile}
\graphicspath{{figures/}}
\newlength\figwidth
\usepackage{url,hyperref}
\usepackage[table,usenames,dvipsnames]{xcolor}
\hypersetup{colorlinks=true, linkcolor=BrickRed,
  urlcolor=blue!50!black, citecolor=blue!50!black}
\usepackage[capitalize]{cleveref}
\crefrangelabelformat{equation}{(#3#1#4)$-$(#5#2#6)}

\newcommand{\reals}{\mathbb{R}}

\newcommand{\VXi}{\VEC{\Xi}}
\newcommand{\VX}{\VEC{X}}

\newcommand{\vn}{\vec{n}}
\newcommand{\vp}{\vec{p}}

\newcommand{\Vp}{\VEC{p}}
\newcommand{\vq}{\vec{q}}
\newcommand{\Vq}{\VEC{q}}

\newcommand{\Vtotal}{V_{\text{tot}}}
\newcommand{\Fmean}{\vec{F}_{\text{av}}}

\begin{document}

\title{Nonequilibrium induced by reservoirs: Physico-mathematical models and numerical tests}

\newcommand\FUBaffiliation{\affiliation{Freie Universit\"{a}t Berlin, Institute of Mathematics, Arnimallee 6, 14195 Berlin, Germany}}

\author{Rupert Klein}
\FUBaffiliation
\author{Roya Ebrahimi Viand}
\FUBaffiliation
\author{Felix H\"{o}fling}
\FUBaffiliation
\affiliation{Zuse Institute Berlin, Takustr. 7, 14195 Berlin, Germany}
\author{Luigi Delle Site}
\email{luigi.dellesite@fu-berlin.de}
\FUBaffiliation

\begin{abstract}
  In a recently proposed computational model of open \fh{molecular} systems out of equilibrium [Ebrahimi Viand \emph{et al.} J.\ Chem.\ Phys.\ \textbf{153}, 101102 (2020)], the action of different reservoirs enters as a linear sum into the Liouville-type evolution equations for the open system's statistics. The \fh{linearity of the coupling} is common to different mathematical models of open systems  and essentially relies on neglecting the
  \fh{feedback} of the system onto the reservoir \fh{due to their interaction}.
  In this paper, we test the range of \fh{applicability} of the \fh{computational} model with \fh{a linear coupling to two different reservoirs, which induces a nonequilibrium situation.}
  \fh{To this end, we studied the density profiles of Lennard-Jones liquids in large thermal gradients using nonequilibrium molecular dynamics simulations with open boundaries.}
  We put in perspective the formulation of an extension of the mathematical model that \fh{can account for} nonlinear effects.
\end{abstract}

\maketitle

\section{Introduction}

Theory and modeling of open systems are becoming increasingly prominent since they allow one to focus on the relevant regions where a process of interest is taking place. The exterior can be instead simplified in the form of thermodynamic reservoirs of particles and energy and is controlled by few macroscopic variables \cite{physrep}. In particular, open molecular systems are of relevance because of their occurrence in \fh{a variety of} current cutting edge technologies, thus they require well-founded numerical algorithms for their efficient and accurate numerical simulation \cite{softmatt}.

In this perspective, physico-mathematical models of open systems represent a guideline protocol for the development of simulation algorithms.
Established models such as the one by Bergmann and Lebowitz \cite{leb1,leb2} (BL) express the combined actions of the reservoirs in the \fh{Liouville-like} equation of the statistical evolution of the open system \fh{by adding the contributions of each single reservoir linearly and independently [cf.\ \cref{liouvext} below].} The linearity for the coupling is a direct consequence of the assumption of impulsive interactions between system and reservoirs, that is each interaction is considered a discrete event in time so that the open system interacts separately in time with each reservoir. The linearity of action is also an assumption in the thermodynamic-based model presented by \citet{jmpnoneq} (GBY). There, the system interacts with different external ``ports'' each of which is a source of energy and mass and mechanical work, and the resulting model is built by adding up the contribution of each port.  Both models are based on a drastic \emph{a priori} simplification of the reservoirs, whose microscopic origin is neglected, thus ruling out the possibility of nonlinear effects in the coupling.

Two of the authors\cite{jmp}, have recently proposed a model, inspired by a simulation protocol for open systems, where the microscopic character of the reservoir is taken into account. In a large system (Universe) the degrees of freedom of the particles of the reservoir are analytically integrated out and an equation for the statistical evolution of the open system is derived. The original derivation considers an open system embedded in a single  homogeneous reservoir, but the extension to more than one reservoir is straightforward and is reported in the appendix.
Also in this model the combined actions of the reservoirs in the equation of statistical evolution of the open system enters as the sum of action of each single reservoir. Differently from the other models reported above, this model is not constructed on an \emph{a priory} choice of \fh{a linear sum of reservoir actions.} \fh{Rather,} the latter originates from the hypotheses of (i) two-body short range interactions between the particles and (ii) of statistical independence of the states of reservoir particle residing close to the open system boundary. As a consequence the coupling between the open system and each reservoir occurs only at the interface regions and thus the contribution of each reservoir is reduced to a surface integral at the interface region.

In a recent work, we have embedded the idea of \fh{adding the actions of independent,} concurrent reservoirs in the Adaptive Resolution Simulation approach (AdResS) \cite{adress1,advcomm,advabbas} and treated the case of an open system interfaced with two distinct and disjoint reservoirs at different temperatures \cite{jcpnoneq}.
The encouraging results of Ref.~\citenum{jcpnoneq} raise the question about the range of validity of the \fh{linear approximation of the reservoir action}.
In this paper,
we test the quality of the numerical approach based on the AdResS technique, which rests solely on \fh{the additivity} of the reservoir contributions. The test consists in simulating a Lennard-Jones (LJ) liquid in an open domain set up such that there is a feedback of the open system onto a sizable part of two attached reservoirs. We compare the results of our model with the results of a reference simulation of the Universe in which  all particles are explicitly treated with all their degrees of freedom, but are thermalized at different temperatures in subregions equivalent to the reservoir domains of our model.
With such a comparison we conclude about the \fh{numerical applicability} of the linear approximation. Surprisingly, for a LJ liquid at thermodynamic and gradient conditions common to a large variety of situations in chemical physics, it is shown that the linear hypothesis holds and nonlinear effects are numerically negligible. This is a \fh{promising insight} in the perspective of developing accurate and efficient simulation algorithms.

While encouraging from the numerical point of view, our conclusions also call for a further development of the mathematical models. The BL and the GBY models by construction cannot implement a boundary response of the reservoir,  instead generalizations to nonlinear and memory effects are within the scope of the other model when less restrictive conditions on the range of particle--particle interactions and on the statistics of reservoir states close to the open system boundary are adopted. For example, it is known rigorously \cite{EngquistMajda1977} that, just as a consequence of multidimensional wave propagation in the reservoir, a nonreflecting acoustic boundary condition \emph{must} entail memory effects. Moreover, when the single- and two-particle statistics involving reservoir particles close to the system boundary are permitted to depend on the state of the open system as a whole, then nonlinear effects will arise in addition as discussed in \cref{sec:perspectives}. Such nonlinear and memory effects of the reservoirs are covered only qualitatively here, while a detailed analysis is left for future work.


\section{Mathematical models of open system}
\label{sec:Comparison}

\subsection{Bergmann-Lebowitz model}

The linear coupling of the open system to distinct reservoirs is the starting point of relevant mathematical models that describe the exchange of matter and energy of a system with its surroundings (see, e.g., Refs.\citenum{leb1,leb2,jmpnoneq} and references therein). For example, in the well-established BL model \cite{leb1,leb2}, the Liouville equation for the \fh{phase space density} $f_n(\VX_{n},t)$ of the open system with $n$ particles assumes {\it a priori} the linear sum of the action of $m$ different reservoirs:
\begin{equation}
  \frac{\partial f_n(t,\VX_{n})}{\partial t} + \{f_n(t,\VX_{n}),H_n(\VX_{n})\}
    = \sum_{r=1}^m I_{n,r}^\text{(BL)}[\VX_n, \{f_{n'}(t)\}] \,,
  \label{liouvext}
\end{equation}
\fh{where $\{\cdot, \cdot\}$ denotes the Poisson bracket, $H_n(\VX_n)$ is the $n$-particle Hamiltonian, and the action of the $r$-th reservoir depends on the family $\{f_{n'}(t)\}_{n'\geq 0}$ of phase space densities at time $t$ and is given by the functional}
\begin{multline}
 I_{n,r}^\text{(BL)}[\VX_n, \{f_{n'}(t)\}] = \sum_{n'=0}^{\infty} \int
  \bigl[ K^{r}_{nn'}(\VX_{n},\VX'_{n'})f_{n'}(\VX'_{n'},t) \\
    - K^{r}_{n'n}(\VX'_{n'},\VX_{n})f_n(t, \VX_{n}) \bigr] d\VX'_{n'} \,.
\end{multline}
Each system--reservoir coupling is assumed to consist of an impulsive interaction formalized by a \fh{Markovian} kernel, $K_{nn'}(\VX'_{n'},\VX_{n})$, i.e., a transition probability per unit time from an $n$-particle (open system) and phase space configuration $\VX_{n}$ to $n'$ particles and phase space configuration $\VX'_{n'}$. The overall global effect resulting from the interactions of the open system with its surrounding is assumed to be linear as expressed by the sum over the $I_{n,r}^\text{(BL)}$ in \cref{liouvext} \fh{and each term being linear in the $f_n$}.
This linearity is actually implicit in the assumption of impulsive \fh{and independent} interactions.
\Citet{leb2} also note that the model of impulsive interaction represents only an asymptotic limit which is not always realized.

\subsection{The thermodynamic model of Gay-Balmaz and Yoshimura}

A thermodynamic perspective to justify a linear coupling is instead employed by Gay-Balmaz and Yoshimura \cite{jmpnoneq}, \fh{using a Lagrangian formulation of the dynamic many-particle system.}
In their model, the system interacts with different ``ports'' each of which is a source of energy and mass and mechanical work that can be injected into or adsorbed from the system. The resulting global model is built by adding up the contribution of each port. Such a modeling approach is justified by the application of the first principle of thermodynamics expressed by a time-dependent energy of the system due to the action of the ports:
\begin{equation}
  \frac{dE}{dt} = \sum_{r=1}^{m} \left(P^\text{ext}_{W,r}+P^\text{ext}_{H,r}+P^\text{ext}_{M,r} \right) \,,
\end{equation}
where $P^{ext}_{W,r}$ is the power corresponding to the work done by the $r$-th reservoir on the system and $P^{ext}_{H,r}$ and $P^{ext}_{M,r}$, respectively, are the power corresponding to the heat and matter transfer from the $r$-th reservoir to the system.

\subsection{Model with marginalization of the degrees of freedom of the reservoir}

\begin{figure}
  \includegraphics[width=\figwidth]{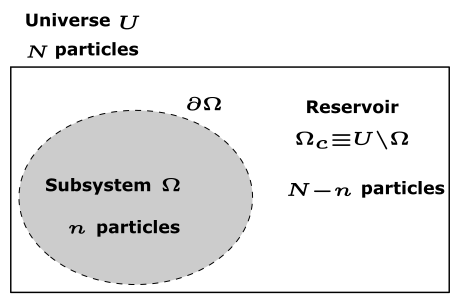}
  \caption{The partitioning of the  ``Universe'' into the open system $\Omega$ and the reservoir $\Omega_{c} \equiv U\setminus\Omega$. The number of particles in the universe is fixed to $N$, but it can fluctuate in the domain $\Omega$ due to exchange with the reservoir.}
  \label{eqfig}
\end{figure}

In this section we report the essential features of the model developed in Ref.~\citenum{jmp}, which are required for the discussion about the linear action of concurrent reservoirs. An extended explanation of the model and its extension to the case of many concurrently acting reservoirs at different thermodynamic conditions are reported in \cref{sec:OpenSystemModel,sec:TwoReservoirs}.
The model considers a large closed system of $N$ particles, the ``Universe'' $U$ (\cref{eqfig}). The Universe is statistically described by its phase space density function $F_{N}(t,\VX^N)$, where $t$ is the time variable and $\VX^N$ are the $6N$-dimensional coordinates in phase space $S^N = \Omega^N \times \mathbb{R}^{3N}$. The time evolution of $F_{N}(t,\VX^N)$ follows the corresponding Liouville equation.
A subsystem $\Omega \subset U$ of the Universe with $n$ particles is described by the probability distribution function, $f_{n}(t,\VX^n)$, obtained by marginalizing $F_{N}(\VX^N)$ w.r.t.\ the $N-n$ particles located in the reservoir $\Omega_c = U \setminus \Omega$:
\begin{equation}
f_n(t, \VX^n) = \binom{N}{n}
  \int\limits_{S_c^{N-n}}
  F_N(t, \VX^n, \VXi_n^{N})\ d\VXi_n^{N} \,;
\end{equation}
\fh{$\VXi_n^{N}$ indicates the degrees of freedom in the reservoir phase space $S_c^{N-n} = \Omega_c^{N-n} \times \mathbb{R}^{3(N-n)}$.
The binomial factor is chosen such that the hierarchy of phase space densities $\{f_n\}_{0\leq n \leq N}$ satisfies} the normalization condition
$
  \sum_{n = 0}^{N} \int_{S^n} f_n(t,\VX^n)\, d\VX^n = 1.
$
The procedure of marginalization is then applied to the Liouville equation of $F_{N}(\VX^N)$, leading to a hierarchy of equations for the $f_{n}(t,\VX^n)$:
\begin{equation}
\frac{\partial f_n}{\partial t}
  + \{f_n, H_n\}
  = \Psi_n + \Phi_{n}^{n+1} \,; \quad 0 \leq n \leq N,
\label{eqliouvsub}
\end{equation}
\fh{where the r.h.s.\ represents the coupling between the system $\Omega$ and the exterior.
Specifically, $\Psi_n = \Psi_n[\VX^n, f_n]$ stems from the forcing of the system particles by the reservoir and $\Phi_{n}^{n+1} = \Phi_{n}^{n+1}[\VX^{n+1}, f_n, f_{n+1}]$ describes the exchange of one particle between the system and the reservoir.
}

\begin{figure}
  \includegraphics[width=\figwidth]{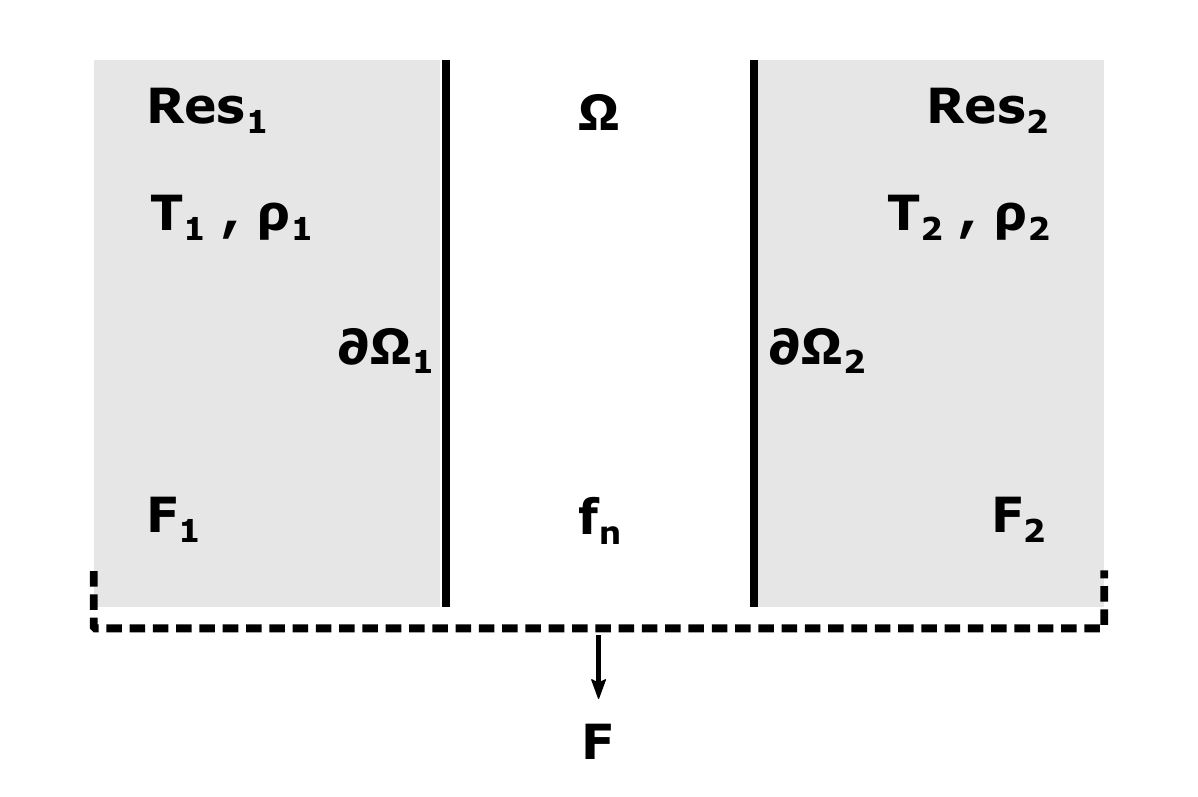}
  \caption{The partitioning of the ``universe'' in two large reservoirs, $\Res_{1}$ and $\Res_{2}$, and the subsystem of interest, $\Omega$.
  The boundary of the open system splits according to $\partial\Omega\equiv \partial \Omega_{1}\cup\partial \Omega_{2}$ into the boundary surface $\partial \Omega_{1}$ between $\Res_{1}$ and $\Omega$
  and the boundary surface $\partial\Omega_{2}$ between $\Res_{2}$ and $\Omega$.
  The two reservoirs are set up at different thermodynamic conditions, e.g., different densities, $\rho_{1}$ and $\rho_{2}$, and different temperatures, $T_{1}$ and $T_{2}$.}
\label{outofeq}
\end{figure}

The derivation above is done under the assumption that the reservoir is thermodynamically uniform.
However, one can imagine that the exterior of $\Omega$ is formed by $m$ disjoint regions at different thermodynamic conditions (see \cref{outofeq} for an example for two regions). Let us assume that the two regions acting as reservoirs are large enough so that we can consider each of them to be in a stationary state within the time scale of observation that we are considering. In such a case, \cref{eqliouvsub} becomes:
\begin{equation}
\label{noneqeq}
\frac{\partial f_n}{\partial t}
  + \{f_n, H_n\}
  = \sum_{r=1}^{m} \left(\Psi_{n,r}+ \Phi_{n,r}^{n+1} \right) ,
\end{equation}
where the sum over $r$ expresses the \fh{additive effect} of the $m$ reservoirs.
The detailed derivation of \cref{noneqeq} along the lines of Ref.\citenum{jmp} is conceptually simple, but involves few specific modifications of the model at the different boundaries of $\Omega$;
a step by step derivation is given in \cref{sec:TwoReservoirs}.
\fh{Most importantly, the contributions $\Psi_n,r$ and $\Phi_{n,r}^{n+1}$ are linear in the $f_n$ and \cref{noneqeq} formally resembles \cref{liouvext}.}

\section{The \protect\AdResS{} setup with the linear \protect\fh{combination} of reservoirs}
\label{sec:NumericalImplementation}

\begin{figure}
\centering
\includegraphics[width=\figwidth]{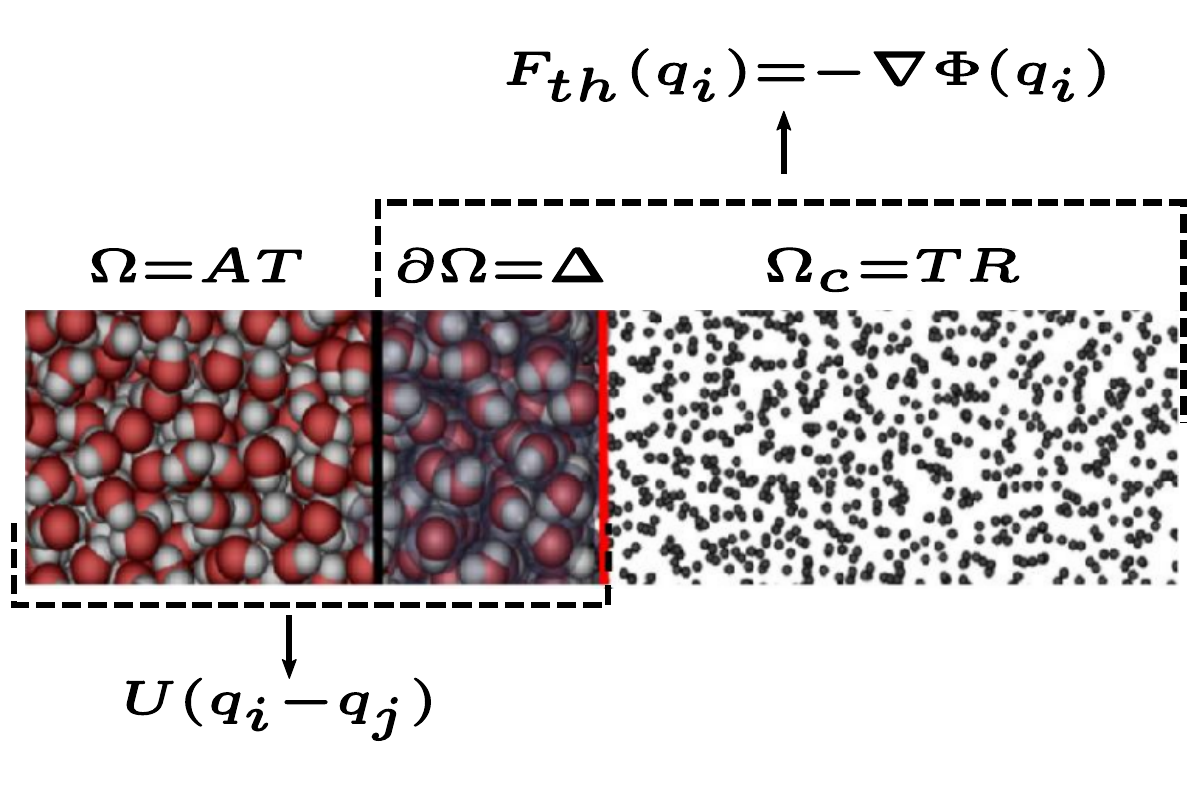}
\caption{The AdResS setup consists of an atomistic region $AT$ and an interface region $\Delta$, both where molecule $i$ interacts with molecule $j$ through the pair potential $U_{ij}=U({\vec q}_{i}-{\vec q}_{j})$. The $\Delta$ region \fh{is interfaced (to the right) with} a large large reservoir TR of noninteracting particles (tracers). In the $\Delta \cup \mathrm{TR}$ region, the thermodynamic force, $F_\text{th}(\vec q_i)$, acts on all particles individually to \fh{enforce} the desired thermodynamic equilibrium.
The correspondence with the mathematical model of open systems is illustrated by identifying each region of AdResS with the equivalent region of \cref{eqfig,outofeq}.}
\label{cartoonadress}
\end{figure}

In a recent work \cite{jcpnoneq}, we have employed the Adaptive Resolution Simulation technique (AdResS) for molecular dynamics \cite{advcomm} to test the concept of a linear combination of reservoir actions on an open system.
The AdResS setup consists of partitioning the simulation box in three regions: the region of interest AT, at full atomistic resolution, the interface region $\Delta$, at full atomistic resolution, but with additional coupling features to the large reservoir, and TR, the large reservoir of noninteracting particles (\cref{cartoonadress}). Particles can freely cross the boundaries between the different regions and automatically acquire the molecular resolution that characterize the region in which they are instantaneously located.

Regarding the coupling conditions, molecules of the AT region interacts with atomistic potentials among themselves and with molecules in $\Delta$, and vice versa, while there is no direct interaction with the tracer particles.
Tracers and molecules in $\Delta$ are subject to an additional one-body force, named thermodynamic force, which acts along the direction $\vec n$ perpendicular to the $\Delta$/TR interface, \fh{$\vec F_\text{th}(\vec q) = F_\text{th}(\vec q)\vec{n}$} for positions $\vec q$.
In essence, this is the coupling condition between the $\Delta$ region and the reservoir TR, \fh{amended by a thermostat in these regions.} As a consequence the total potential energy reads: $U_\mathrm{tot}= U_\mathrm{tot}^\mathrm{AT}
+ \sum_{\vec q_j\in \Delta \cup \mathrm{TR}}\phi_\text{th}(\vec q_j)$ with \fh{the potential $\phi_\text{th}(\vec q)$ such that
$\vec F_\text{th}(\vec q) = -\nabla \phi_\text{th}(\vec q)$ and $\phi_\text{th}(\vec q) = 0$ in the AT region, $\vec q\in \mathrm{AT}$.}
The thermodynamic force is derived by basic principles of statistical mechanics; in essence, of relevance for this paper, it assures that the particle density in the atomistic region is equal to a value of reference. As it is shown in Refs.\citenum{jcpsimon,prl2012,prx,advabbas} the constraint on the particle density in AdResS implies the equilibrium of the atomistic region w.r.t. conditions of reference of a fully atomistic simulation.

\begin{figure}
\centering
\includegraphics[width=\figwidth]{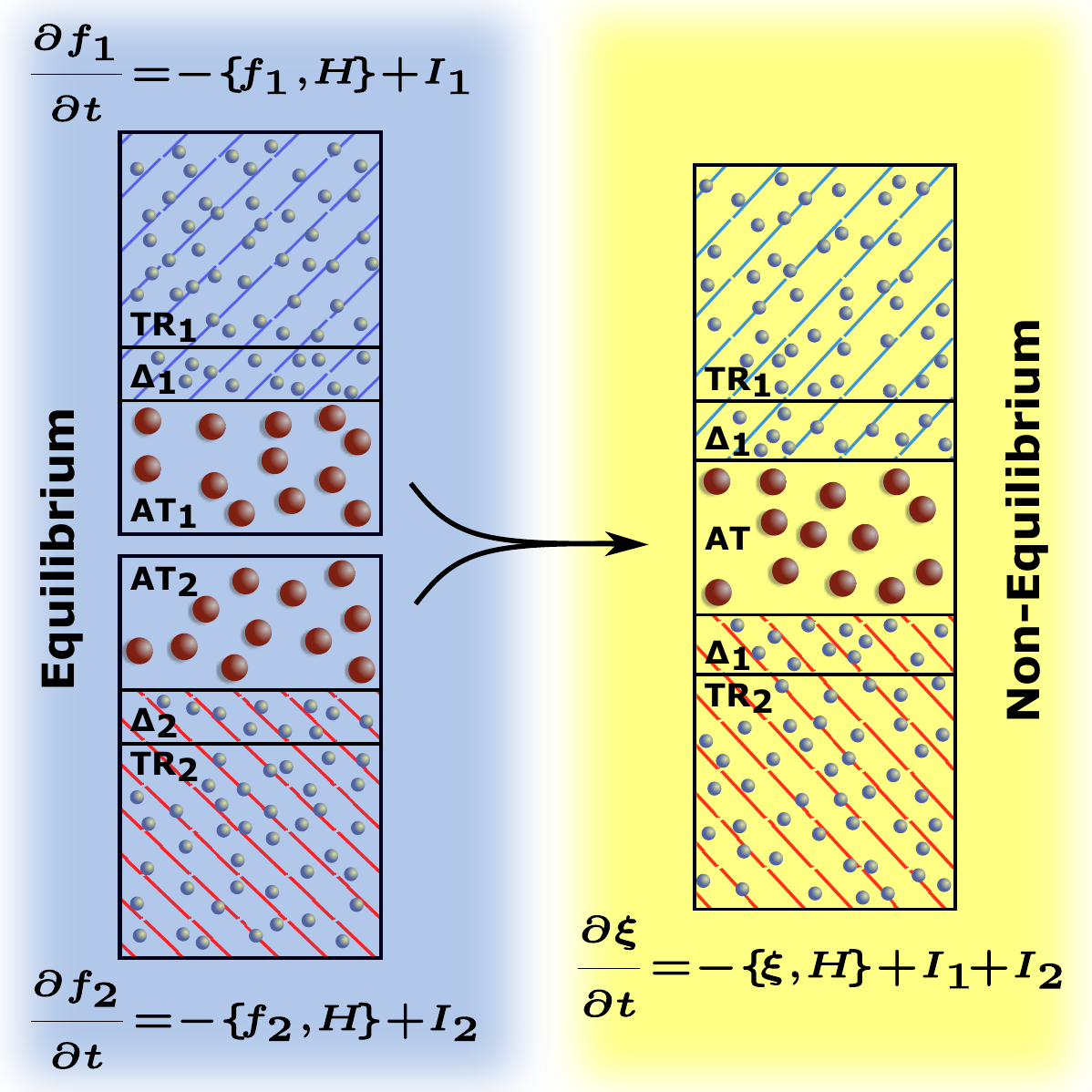}
\caption{Schematic illustration of the simulation of a thermal gradient in the AdResS setup and its correspondence to the mathematical models. First, the open system is equilibrated at the thermodynamic condition of each reservoir (left).
In AdResS, this corresponds to running the equilibration procedure twice to determine the thermodynamic force $F_{\text{th},1}(\vec q)$ and $F_{\text{th},2}(\vec q)$ separately. Once the system is in contact with two different reservoirs (right), then the mathematical models \fh{predict} a linear action of the reservoirs \fh{as is apparent from the r.h.s.\ $I_1+I_2$ of the extended Liouville equation [cf.\ \cref{liouvext,noneqeq}]}.
The reservoir coupling terms $I_r$ ($r=1,2$) translate in AdResS to the combined action of $F_{\text{th},1}(\vec q)$ and a thermostat that maintains the temperature at $T_{1}$ in the region $\Delta_{1} \cup \text{TR}_1$ and analogously for the second reservoir; in this sense, $I_r$ is a function of $F_{\text{th},r}(\vec q)$ and the thermostat at $T_{r}$.}
\label{noneqfig}
\end{figure}

The setup of AdResS \fh{resembles} the partitioning employed in the mathematical models of open system and, in particular, it is very well suited for a numerical test of the idea of a linear action of reservoirs. In fact, \fh{in AdResS one can implement} a setup as that of \cref{outofeq}, where the action of the two distinct reservoirs, $\Res_1$ and $\Res_2$, is encoded in two distinct coupling conditions at the corresponding interfaces. The coupling terms, which correspond to the thermodynamic forces, are calculated separately; that is, the system first interacts only in the presence of $\Res_{1}$, which is at temperature $T_{1}$ and density $\rho_{1}$, and one obtains the thermodynamic force needed at the interface with $\Res_{1}$. Next, the system interacts only in the presence of $\Res_{2}$, at $T=T_{2}$ and $\rho=\rho_{2}$, and one obtains the thermodynamic force needed at the interface with $\Res_{2}$.
A nonequilibrium situation is then achieved by running a simulation setup with distinct thermodynamic forces applied in the corresponding interface regions (\cref{noneqfig}).

\section{Numerical tests of the linear approximation of the reservoir action}

The AdResS approach to open systems out of equilibrium was applied to simulate a LJ liquid in a temperature gradient \cite{jcpnoneq}. The results of this earlier study showed that indeed the model accurately reproduces data from fully atomistic reference simulations in the presence of a thermal gradient. Specifically, an isobaric setup was employed, that is the temperature gradient is applied at constant pressure \fh{by choosing reservoir densities along an isobar for prescribed reservoir temperatures.}
Here we go further and consider situations where the idea of linearity is pushed to its edge of validity. To this aim, we numerically test what happens in the atomistic region of interest when at the interface regions one has a feedback from the rest of the system. We performed nonequilibrium simulations (i) along an isobar with increasing temperature gradients, whose largest value exceeds the one of our earlier work \cite{jcpnoneq} by a factor of~3 and (ii) in an isochoric setup, that is, the thermodynamic forces are calculated \fh{at the same density $\bar\rho=\rho_{1}=\rho_{2}$, but different reservoir temperatures $T_1 < T_2$.}
\fh{We note that the reservoir states are chosen in the liquid phase and are close to the liquid--vapour binodal curve; here, the LJ fluid is almost incompressible and is characterized by low pressure.}
The technical details of the simulations are given in \cref{sec:technicalsection}.

\begin{figure*}
\includegraphics[width=\figwidth]{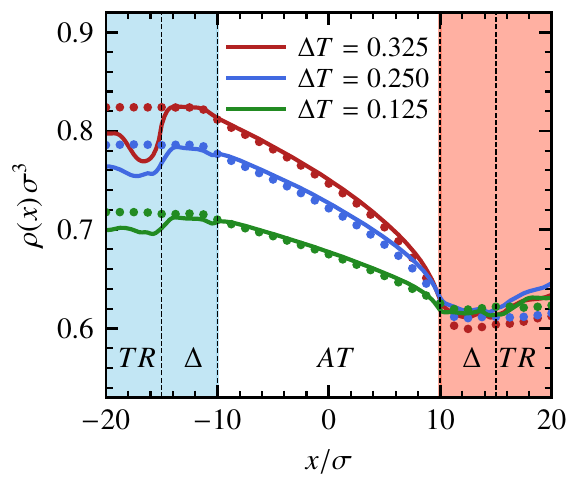}%
\caption{Density profiles of a LJ liquid in different thermal gradients, simulated with the isobaric setups of nonequilibrium AdResS (solid lines). The temperature difference $\Delta T$ between the hot (red) and cold (blue) reservoirs increases from \fh{bottom to top} as indicated in the \fh{legend}. The temperature and density of the hot reservoir are kept fixed at $T_\text{hot} = 0.95 \epsilon/\kB$ and $\rho_\text{hot} = 0.622\sigma^{-3}$, respectively, while the state points of the cold reservoir are chosen along the corresponding isobar. Here, $\epsilon$ and $\sigma$ refer to the parameters of the LJ potential, see \cref{sec:technicalsection}. Reference results from full atomistic simulations are given by disc-shaped symbols. Only the parts of the TR regions close to the coupling boundary are shown.
}
\label{highgrad}
\end{figure*}

\fh{For the isobaric setup, case (i), the density profiles across the simulation box obtained for different thermal gradients
(\cref{highgrad}) follow closely the results of the corresponding, fully atomistic reference simulations (which involves the atomistic simulation of a huge reservoir), in particular in the region AT of interest.
The highly satisfactory agreement is qualitatively similar for all temperature gradients investigated, despite $\Delta T:=T_2 - T_1$ increasing from 14\% to 40\% relative to the respective mean temperature, $\bar T = (T_1 + T_2)/2$.}
In the $\Delta \cup \mathrm{TR}$ there are noticable effects due to a feedback of the system onto the reservoir, without repercussions on the region of interest.

\begin{figure}
\centering
\includegraphics[width=\figwidth]{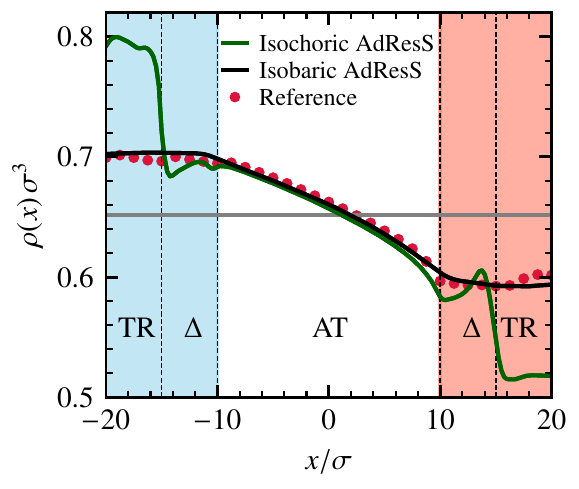}
\caption{Density profiles of a LJ liquid in a fixed thermal gradient ($\Delta T=0.125\epsilon/\kB$) simulated with the isochoric (green line) and isobaric (black line) setups of nonequilibrium AdResS; reference data from full atomistic simulations are given by red symbols.
\fh{The temperature of the hot reservoir is $T_\text{hot} = 0.975 \epsilon/\kB$, and the mean particle density, averaged over the whole setup, is $\bar\rho\approx 0.65\sigma^{-3}$ (grey line).}
Only parts of the TR regions are shown.
}
\label{noneqisoc}
\end{figure}

For the isochoric nonequilibrium setup, case (ii), \fh{one expects that the thermal gradient induces a pressure gradient in the AT region. Interestingly, our simulation results (\cref{noneqisoc}) instead exhibit a density gradient that closely follows the one of the isobaric AdResS setup and the full atomistic reference. A possible explanation is that the pressure gradient induces a mass flux, which builds up a density gradient until the pressure differences are compensated.
The interpretation is corroborated by the large density difference between the two TR regions, reflecting tracers that were moved in excess from the hot to the cold reservoir.
This mechanism introduces a feedback on the two reservoirs and yields a sizeable shift of their intended densities.
Nevertheless, also the isochoric AdResS setup provides a meaningful (and simpler) approach to nonequilibrium simulations.
}
It shall be added that the temperature profiles of all AdResS setups considered in \cref{highgrad,noneqisoc} agree extremely well with the simulations of reference and are not shown here, see Ref.~\citenum{jcpnoneq} for an example.

\section{Theoretical Perspectives}
\label{sec:perspectives}

From the physical point of view, both the BL and GBY models rely on neglecting any feedback of the system onto the reservoirs. The former model also predicts a linear combination of the action of different reservoirs as long as the \fh{condition} of short range pair interactions and the explicit assumption of statistical independence of reservoir states from the open system states apply. However, such constraints can be relaxed and nonlinear and memory effects can be introduced into the derivation. Both types of effects are expected in general, as the following examples demonstrate:
\begin{itemize}
\item Memory effects: We are dealing with thermodynamically compressible systems supporting sound waves. It is known from the theory of fluid dynamics that the modelling of acoustic waves that exit an open domain without reflection at its boundary \fh{demand that} one keeps track of the system's history to properly capture the wave dynamics in the reservoir\cite{EngquistMajda1977}.

\item Nonlinearities: When the open system is at a higher pressure than the reservoir(s) initially, then the velocity statistics of outer particles near the open system boundary is clearly biased towards the outward pointing normal. This effect could be modelled by introducing single- and two-particle statistics $f^\circ_{n,1}$ and $f^\circ_{n,2}$ that depend explicitly on the particle number $n$ of the open system and by coupling these functions to thermodynamic averages of the entire hierarchy of \fh{phase space densities} $f_{\nu}$, $\nu \in \{0, ..., N\}$. This ansatz will \fh{render} the terms $\Psi_n$ and $\Phi_{n}^{n+1}$ in \cref{noneqeq} \fh{nonlinear functions of the $f_n$}. If nonlinear effects are to arise in the presence of long-range interactions, then the model would similarly be able to account for these.

\item \fh{Long-range correlations: Another possibility is that, in the vicinity of a critical point, the correlation length of the fluid is of the order of the linear size of the open system.
In this case, the presence of the open boundaries (i.e., the finite size of the subsystem) unavoidably modifies the observed fluid properties, e.g., its thermodynamics \cite{Schnell:CPL2011} and the strength of local density fluctuations \cite{finiteslab:JCP2020, Pathania:ATS2021}, and one anticipates a direct correlation between the reservoirs.
Truncating such a correlation spectrum is known to give rise to critical Casimir forces: an effective, non-additive interaction between the boundaries, which was found for (binary) fluids between solid and periodic boundaries near criticality \cite{Hertlein:N2008,Paladugu:NC2016,Puosi:PRE2016}, but also transiently after a temperature quench \cite{Rohwer:PRL2017}. Capturing such an extreme situation by a model of open systems would require that detailed information about the reservoir states is kept, with the degree of detail depending on the observables of interest.}
\end{itemize}

Yet, the model of Ref.~\citenum{jmp} tells something more. Even within the assumption of short-range interactions, one still needs basic information about the reservoirs at the interface, e.g., one- and two-particle distributions. These latter are assumed to be stationary, under the approximation that the reservoir, due to its size, fully controls the particle distribution at the surface. Such an approximation is no more valid if the range of interaction is extended because the system itself starts to significantly influence the distribution of particles near the boundary and a generally nonlinear feedback between the reservoirs is to be expected.

\section{Conclusions}
\label{sec:Conclusion}

We have discussed models of open systems \fh{that rely on} the linear combination of the actions of several reservoirs. From the general discussion emerges the necessity of considering the scenario in which nonlinear effects arise once the hypotheses of short range potentials and statistical independence of the reservoirs becomes less strict. In such a perspective, a model recently proposed by two of the authors offers the possibility of automatically including the feedbacks between the open system and the attached reservoirs by generalizing  the pertinent models of the reservoir statistics at each interface. Such a possibility is ruled out in the other models of an open system covered here. An analytic derivation of such a generalization is, however, left to future research.
Instead we have focused on the numerical consequences of such nonlinear effects and reported a numerical test where the reference simulation automatically includes nonlinear effects while the simulation \fh{mimicking the mathematical model of \cref{noneqeq}} does not include \fh{such a nonlinearity}. The results show that even under such conditions, the approximation of a linear combination of actions yields numerically satisfying results for a range of systems of interest.
\fh{This is good news} for the numerical simulations, \fh{in particular, it removes the need to fine-tune the reservoir states of the isobaric setup. At the same time, our findings} also call for an extension of the concepts and its numerical counterpart to nonlinear effects.

\acknowledgments{This research has been funded by Deutsche Forschungsgemeinschaft (DFG) through grant CRC 1114 ``Scaling Cascade in Complex Systems,'' Project Number 235221301, Project C01 ``Adaptive coupling of scales in molecular dynamics and beyond to fluid dynamics.''}

\appendix

\section{Liouville-type equations for an open system}
\label{sec:OpenSystemModel}

In this appendix, we summarize the key steps of the modeling procedure adopted in Ref.\citenum{jmp} {for an open system embedded in a single uniform reservoir}. This serves as basis for the extension of the model to several independently acting reservoirs in \cref{sec:TwoReservoirs}.


\subsection{Topological definition of an open system}

Let us consider the open system schematically illustrated in \cref{eqfig}. The total system, here called ``Universe'', is characterized by the number $N$ of particles (fixed) and a spatial domain $U$.
A subdomain $\Omega \subset U$ defines the open system, which may contain any number $n \in \{0, ..., N\}$ of particles, while the region of the reservoir corresponds to the complement $U \setminus \Omega\equiv \Omega_{c}$ with $N-n$ particles. The phase space of the $N$ particles in $U$ is $S^{N}=U^{N} \times \mathbb{R}^{3N}$ (positions in $U^{N}$, momenta in $\mathbb{R}^{3N}$), the phase space of the open system is in case of $n$ particles is $S^{n}=\Omega^{n} \times \mathbb{R}^{3n}$, and the phase space of the reservoir is $S_{c}^{(N-n)}=\Omega_{c}^{(N-n)} \times \mathbb{R}^{3(N-n)}$.


\subsection{Relevant quantities that characterize $U$}
\label{ssec:RelevantQuantitiesUniverse}

The Universe is characterized by the Hamiltonian
\begin{equation}
H_{N}({\bf q}^{N},{\bf p}^{N})
= E_{\rm kin}({\bf p}^{N}) + V_{\rm tot}({\bf q}^{N})
\equiv \sum_{i=1}^{N} \frac{(\vp_i)^{2}}{2M} 
     + \frac{1}{2}  \mathop{\sum_{i,j=1}^{N}}\limits_{j\not=i} V(\vq_j-\vq_i) 
     \label{UH}
\end{equation}
where
$
\left({\bf q}^{N},{\bf p}^{N},\right)
= (\vec{q}_1, ..., \vec{q}_N, \vec{p}_1, ..., \vec{p}_N)
$
and $(\vec{q}_i, \vec{p}_i)$ are the position and momentum of the $i$-th particle, respectively, $M$ is the particle mass, assumed to be the same for all particles here, and $V$ is the two-body interaction potential (as typical for, e.g., molecular dynamics simulations). The statistical mechanics description of the system in phase space is achieved through its probability density defined as
\begin{equation}
 F_N : \reals^+ \times S^N \to \reals \,, \quad
 (t, \VX^N) \mapsto F_N(t,\VX^N) \,,
\label{UF}
\end{equation}
with normalization $\int_{S^N} F_N\, d\VX^N = 1$. The probability density of the Universe is subject to the transport equation of the phase space density (Liouville equation)
\begin{equation}
\frac{\partial F_N}{\partial t}= \sum\limits_{i=1}^N 
\left[
\nabla_{\vec{q}_i} \cdot \left(\vec{v}_i F_N\right)
+
\nabla_{\vec{p}_i} \cdot \left(-\nabla_{q_i}\Vtotal(\Vq^N) F_N\right)
\right] \equiv -\{F_N, H_N\}
\label{UL}
\end{equation}
where the r.h.s.\ is a Poisson bracket and $\vec{v}_i = \vec{p_i}/M$ the velocity of the $i$-th particle.


\subsection{Relevant quantities that characterize $\Omega$}
\label{ssec:RelevantQuantitiesOpenSystem}

The open system $\Omega$ \fh{containing $n$ particles} is characterized by the Hamiltonian
\begin{equation}
H_n 
= \sum_{i=1}^{n} \frac{(\vp_i)^2}{2M} 
+ \frac{1}{2} \mathop{\sum_{i,j = 1}^{n}}\limits_{j\not=i} V(\vq_j-\vq_i) \,;\qquad (\vq_i, \vq_j \in \Omega)\,.
\label{OH}
\end{equation}
The statistical mechanics description of the open system in phase space is given by the collection of all its $n$-particle probability densities
\begin{equation}
  f_n: \reals^+ \times S^n  \to \reals \,, \quad
  (t, \VX^n) \mapsto f_n(t, \VX^n) \,,
  \label{Of}
\end{equation}
for $n \in \{0, ..., N\}$. Consistent with the fact that $\Omega$ is a subsystem of $U$, $f_{n}$ is explicitly given by \cite{jmp}:
\begin{equation}
  f_n(t, \VX^n)
= \binom{N}{n}
  \int\limits_{(S_{c})^{N-n}}
  F_N(t, \VX^n, \VXi_n^{N})\ d\VXi_n^{N}
  \label{fdef}
\end{equation}
with $\VXi_n^{N}  \equiv [X_{n+1},.....{X_{N}}]$ \fh{collecting the reservoir's degrees of freedom,}
$X_{i}=(\vq_i,\vp_i) \in S_{c}$, and the normalization condition:
$\sum_{n = 0}^{N} \, \int_{S^n} f_n(t,\VX^n)\,d\VX^n = 1$.
\fh{The binomial factor counts the number of ways to pick $n$ particles out of $N$.}


\subsection{Derivation of a Liouville-like equation for $\Omega$}
\label{ssec:LiouvillEqnDerivation}

The starting point is \cref{UL} and the strategy to achieve a Liouville-like equation for $f_{n}$ consists in the marginalization of \cref{UL} w.r.t.\ the degrees of freedom of the $N-n$ particles in $\Omega_{c}$. The procedure can be schematized by the following two steps
\begin{enumerate}[\bfseries I)]
\item Marginalization of the term $\sum_{i=1}^{N}\nabla_{\vec{p}_i} \cdot \left(-\nabla_{q_i}\Vtotal(\Vq^N) F_N\right)$. Since $\Vtotal(\Vq^N)$ \fh{is a sum over all index pairs $1\leq i,j\leq N$,} one needs to analyze three specific situations:
\begin{align}
  (a) \quad & i,j\in\Omega_{c}; & (b) \quad & i,j\in\Omega; & (c) \quad & i\in\Omega,j\in\Omega_{c}.
  \label{stepI}
\end{align}

\item Marginalization of the term $\sum\limits_{i=1}^N \nabla_{\vec{q}_i} \cdot \left(\vec{v}_i F_N\right)$, where two specific situations need to be distinguished:
\begin{equation}
 (a) \quad i \in\Omega \qquad \text{and} \qquad (b) \quad  i\in\Omega_{c} \,.
 \label{stepII}
\end{equation}

\end{enumerate}
%


\subsubsection*{Results of Step I}

\begin{enumerate}[(a)]
   \item \fh{For all particle indices $i$ in the reservoir,} $n+1\le i\le N$, \fh{Gauss' theorem implies:}
   \begin{multline}
   \int\limits_{B_r(0)} 
     \nabla_{\vec{p}_i}\cdot\bigl(\nabla_{q_i}\Vtotal(\Vq^N)F_N(t, \Vq^N, \Vp^N)\bigr) d^3 p_i \\
     = \int\limits_{\partial B_r(0)} \vec{n}\cdot\left(\nabla_{q_i}\Vtotal(\Vq^N)\right) F_N(t, \Vq^N, \Vp^N) \,d\sigma_{p_i} \to 0
   \label{stepIa}
   \end{multline}
   as $r \to \infty$,
   where $B_r(0)$ is the sphere of radius $r$ in momentum space centered at the origin
   \fh{and $\vec n$ the surface normal on $\partial \Omega$ pointing outwards}.
   It is assumed that
   $F_{N}$ decays sufficiently rapidly for large $|\vec{p}_i|$ for the boundary integral to vanish in the limit. 
   This is certainly true, e.g., for the Boltzmann distribution \fh{of the momenta,}
   $F_N \propto \exp\bigl(-(\vec{p}_{i})^{2}/2MkT\bigr)$.
Actually, it is sufficient that $F_N$ decays to zero at all for large momenta and this is a consequence of $F_N$ being a probability density and thus integrable.
     
   \item \fh{If both particles $i,j$ are in the open system $\Omega$,} $1 \leq i,j \leq n$, \fh{marginalization over the reservoir yields:}
   \begin{multline}
   \binom{N}{n} \!\!\!  \int\limits_{(S_{c})^{N-n}} \!\!\!
\nabla_{\vec{p}_i}\cdot\Bigl(\nabla_{q_i}V(\vq_i-\vq_j)F_N(t, \VX^n, \VXi_n^{N})\Bigr) d\Xi_n^N \\
= \nabla_{\vec{p}_i}\cdot\Bigl(\nabla_{q_i}V(\vq_i-\vq_j)f_n(t, \VX^n)\Bigr).
   \label{stepIb}
   \end{multline}

\item \fh{If particle $i$ is in the open system,} $1\le i\le n$, \fh{but particle $j$ is in the reservoir,} $n+1\le j\le N$, \fh{then choosing $i=n$ and $j=n+1$ for the ease of notation, one finds:}
  \begin{multline}
  \binom{N}{n} \int\limits_{S_{c}}\int\limits_{(S_{c})^{N-n-1}}
    \nabla_{\vec{p}_i}\cdot
  \Bigl(-\nabla_{q_i}V(\vq_i-\vq_j)F_N(t, \VX^{n-1,}, X_i, (\vq_j,\vp_j), \VXi_{n+1}^{N})\Bigr) {{d\VXi_{n+1}^{N}}} dp_j dq_j
      \\
     = \nabla_{\vec{p}_i}\cdot\Bigl(\Fmean(\vq_i) f_n(t, \VX^{n-1}, X_{i})\Bigr)
  \label{stepIc}
  \end{multline}
\end{enumerate}
with 
\begin{equation}\label{eq:f2}
\Fmean(\vq_i) = - \int\limits_{S_{c}} \nabla_{\vq_i}V(\vq_i- \vq_j) f^\circ_2(X_j | X_i) d X_j
\end{equation}
 denoting the mean-field force exerted by the outer particles onto the $i$-th inner particle under the assumptions that
\begin{enumerate}[(1)]

  \item\label{it:Ass1}  pair interactions $V({\bf q}_{i}-{\bf q}_{j})$ are short-ranged so that pair interactions are relevant only close to the open system's boundary,

  \item\label{it:Ass2} the probability density of finding $n$ particles in states $\left(\VX^{n-1}, X_i\right) \in S^n$ and one other outer particle in $X_j$, \fh{given by marginalization over $\VXi_{n+1}^N$,} can be factorized as
  \begin{equation}
       \fh{\binom{N}{n} \int\limits_{(S_{c})^{N-n-1}}
    F_N(t, \VX^{n-1,}, X_i, X_j, \VXi_{n+1}^{N}) {{d\VXi_{n+1}^{N}}}} \approx
     f^{\circ}_2(X_j | X_i) f_n\left(t, \VX^{n-1}, X_i\right).
  \end{equation}

  \item\label{it:Ass3} $f^{\circ}_2(X_{\text{out}} | X_{\text{in}})$ is a known or modelled conditional distribution for joint appearances of an outer particle given the state of an inner one.
  
\end{enumerate}
{Here we consider assumption \ref{it:Ass1} as a physical necessity for assumptions \ref{it:Ass2} and \ref{it:Ass3} to be justifiable in the first place, while the latter two encode the more general assumption that the statistics of the reservoir is independent of the instantaneous state of the open system for the present purposes.}


\subsubsection*{Results of Step II}

\begin{enumerate}[(a)]
  \item \fh{For all particles $i$ in $\Omega$,} $i \in \{1,\dots, n\}$, it holds
    \begin{equation}
    \binom{N}{n} \int\limits_{(S_{c})^{N-n}}
    \nabla_{\vec{q}_i} \cdot \left(\vec{v}_i F_N(t, \VX^n, \VXi_n^N)\right) d\VXi_n^N=\nabla_{\vec{q}_i} \cdot \left(\vec{v}_i f_n\right).
    \label{stepIIa}
    \end{equation}
  \item \fh{In the reservoir,} $n+1 \leq i \leq N$, one of the integrals will be over $\Xi_i \in S_{c}$ which, after summing over the respective terms and utilizing the indistinguishability of the particles, leads to \fh{an integral over the boundary $\partial\Omega$ of the open system:}
   \begin{multline}
   \binom{N}{n} (N-n)
    \int\limits_{S_{c}} 
    \int\limits_{(S_{c})^{N-n-1}}
    \nabla_{\vec{q}_i} \cdot \left(\vec{v}_i \,F_N(t, \VX^n, (\vq_i,\vp_i), \VXi_{n+1}^{N}\right)
    \, d\VXi_{n+1}^N \ d \Xi_i
        \\
    = - (n+1)
        \int\limits_{\partial \Omega} 
          \int\limits_{\reals^3}
          \left(\vec{v}_i \cdot \vn\right) \ {\widehat f}_{n+1}(t, \VX^n, (\vq_i,\vp_i)) \,d^3 p_i d\sigma_i\,,
    \label{stepIIb}
   \end{multline}
   \fh{where we employed the identity $\binom{N}{n}(N-n) = \binom{N}{n+1}(n+1)$ and the notation assumes $i=n+1$.}
\end{enumerate}
Here, guided by the theory of characteristics, we distinguish the relevant forms of ${\widehat f}_{n+1}$ for outgoing and incoming particles as follows: Under the assumption of statistical independence of the reservoir particle states from those of the inner particles, we have
\begin{equation}\label{eq:OuterF}
{\widehat f}_{n+1} = 
\begin{cases}
f_{n+1}
  & \left(\vec{v}_i \cdot \vn > 0\right) \,,
    \\
f_{n}f_1^{\circ}
  & \left(\vec{v}_i \cdot \vn < 0\right) \,,
\end{cases}
\end{equation}
where $f_1^{\circ}$ is the single particle (equilibrium) density of the reservoir. Alternatively, assuming a grand canonical (GC) distribution for state space trajectories that enter the open system from outside one could \fh{write down} as a plausible model:
\begin{equation}\label{eq:OuterFGC}
{\widehat f}_{n+1} = 
\begin{cases}
f_{n+1} 
  & \left(\vec{v}_i \cdot \vn > 0\right) \,,
    \\
f_{n+1}^{\text{GC}}
  & \left(\vec{v}_i \cdot \vn < 0\right) \,.
\end{cases}
\end{equation}
Note that we do not intend to promote the closure assumptions regarding the reservoir statistics introduced above (through the functions $f^\circ_1$ and $f^\circ_2$) as being optimal or preferable over alternative formulations. Instead, these closures are meant to be placeholders that highlight the principal necessity of explicitly formulating assumptions on the reservoir behavior in the context of the present derivations.


\subsubsection*{Final Equation}

Combining the results of Steps I and II, one obtains a hierarchy of Liouville-type equations for $f_{n}$:
\begin{equation}
\frac{\partial f_n}{\partial t} 
  + \fh{\{f_n, H_n\}}
= \Psi_n + \Phi_{n}^{n+1}
\label{OL}
\end{equation}
with \fh{boundary terms on the r.h.s.\ describing the action of the reservoir, namely the interaction term due to a mean-field forcing by the reservoir particles}
\begin{equation}
\Psi_n[\VX^n, f_n]
  =
    - \sum\limits_{i=1}^{n} \nabla_{\vec{p}_i}\cdot\bigl(\Fmean(\vq_i) f_n(t, \VX^{i-1}, X_{i}, \VX_{i}^{n-i})\bigr) \,,
\end{equation}
\fh{and an exchange term due to particles entering and leaving the domain $\Omega$:}
\begin{multline}
\Phi_{n}^{n+1}[\VX^n, f_n, f_{n+1}]
  = \\
    (n+1) \int\limits_{\partial \Omega}
          \int\limits_{(\vec{p}_i \cdot \vn) > 0}
            \hspace{-5pt}
            (\vec{v}_i \cdot \vn) \
            \bigl( {f}_{n+1}\left(t, \VX^{n},  (\vq_i,\vp_i)\right)
                 - {f}_{n}\left(t, \VX^n\right){f}_{1}^{\circ}\left(\vq_i,-\vp_i\right)
            \bigr)
  d^3 p_i d\sigma_i \,.
  \label{defquant}
\end{multline}


\section{The case of two distinct reservoirs at different thermodynamic conditions}
\label{sec:TwoReservoirs}

Let us consider a prototype situation as that illustrated in \cref{outofeq} where $\Omega$ is a region that separates the Universe in two distinct (large) reservoirs, ($\Res_{1}$ and $\Res_{2}$), and, as anticipated before, we assume that the two reservoirs are in stationary thermodynamic conditions in the time scale considered. Straightforward physical considerations lead to the conclusion that $\Omega$ has a spatially asymmetric exchange with the Universe and may thus possess a nonequilibrium stationary state. Formally one can proceed as for the case of a single reservoir and derive an equation for $f_{n}$ in this situation. {The total probability distribution function $F_{N}$ describes the entire Universe including the thermodynamic states of $\Res_{1}$ and $\Res_{2}$}. From $F_{N}$, by marginalizing w.r.t. $N-n$ degrees of freedom of particles in $\Omega_{c}$, one obtains the distribution function $f_{n}$ of the open system. Furthermore, the Liouville equation for $F_{N}$ applies as before, and thus by marginalizing the Liouville equation for $F_{N}$ w.r.t. the degrees of freedom of the particles in $\Omega_{c}$ one would obtain the corresponding Liouville-type equation for $f_{n}$ ($n \in \{0, ..., N\}$) as for the case of one reservoir of Ref.\citenum{jmp}. This means an analytic derivation of the conditions of nonequilibrium induced by the concurrent action of the two reservoirs. In the sections below we will follow the marginalization procedure adopted in the previous section and adapted to the setup illustrated in \cref{outofeq}


\subsection*{Step I revised with $\Res_{1}$ and $\Res_{2}$}

\begin{enumerate}[(a)]
\item \fh{Again, the boundary integral in \cref{stepIa} will vanish in the limit of the radius of the ball tending to infinity because  $F_N$ will decay rapidly for large momenta, i.e., one can expect a decay as $\exp[-(\vec{p}_i)^{2}/2MkT_{1}]$ in $\Res_{1}$ and  $\exp[-(\vec{p}_i)^{2}/2MkT_{2}]$ for $\Res_{2}$ (assuming they are both much larger than $\Omega$). As said before, it is sufficient that $F_N$ is a probability density and thus integrable.}

\item \fh{If both particles $i,j$ of the pair are inside of $\Omega$, nothing changes. In particular,}
the marginalization w.r.t.\ the particles outside [\cref{stepIb}] implies that the whole information about the particles of the reservoirs is integrated out.

\item Here emerges the first substantial difference. \fh{In the case that particle $i$ is inside of $\Omega$ and particle $j$ in one of the reservoirs, \cref{stepIc} remains formally the same, but the calculation of the mean force changes [\cref{eq:f2}].}
One needs to carefully consider the dependency of the pair's potential energy on the position of the particle in each of the two distinct reservoirs.
The \fh{modified} expression of $\Fmean(\vq_i)$ carries the fact that the boundary with $\Res_1$ has different thermodynamic and statistical mechanics properties than the boundary with $\Res_2$, {depending on the specific subdomain of $S_{c}$ over which the integration in the variable $X_{j}$ is carried out}. This implies that the assumed probability density of finding $n$ particles in states $\left(\VX^{n-1}, X_i\right) \in S^n$ and one other outer particle in $X_j$ is now given by $f_n\left(\VX^{n-1}, X_i\right) f^{\circ, R1}_2(X_j | X_i)$ if $X_j\in \Res_{1}$ and  by $f_n\left(\VX^{n-1}, X_i\right) f^{\circ,R2}_2(X_j | X_i)$ if $X_j \in \Res_{2}$.
The integration of $X_{j}$ over the whole \fh{$S_{c} = \Res_{1} \cup \Res_{2}$ splits into} the sum of two integrals \fh{over the domains $\Res_{1}$ and $\Res_{2}$, respectively:}
\begin{equation}\label{eq:meanforce}
\Fmean(\vq_i) 
= 
- \int\limits_{\Res_{1}}
  \nabla_{\vq_i}V(\vq_i- \vq_j) f^{\circ,R1}_2(X_j | X_i) \,d X_j
- \int\limits_{\Res_{2}}
  \nabla_{\vq_i}V(\vq_i- \vq_j) f^{\circ,R2}_2(X_j | X_i) \,d X_j \,.
\end{equation}
\fh{Specifically, the assumption that the statistics of both reservoirs are independent of each other yields the additive form}
$\Fmean(\vq_i) = \Fmean^{R1}(\vq_i)+\Fmean^{R2}(\vq_i)$.
\end{enumerate}


\subsection*{Step II revised with  $\Res_{1}$ and $\Res_{2}$}
\begin{enumerate}[(1)]
  \item \fh{Similarly as in step I,} for the particles $i \in \{1, \dots , n\}$ inside of the domain $\Omega$ \fh{the terms in \cref{stepIIa}} remain the same because the marginalization w.r.t.\ the particles outside implies that any information about the reservoirs is integrated out.

  \item Otherwise, for $n+1 \leq i \leq N$, the effects of the two distinct reservoirs \fh{entering \cref{stepIIb}} clearly emerges in the definition of ${\widehat f}$ [\cref{eq:OuterF}], because one needs to define ${\widehat f}$ \fh{differently on the two reservoirs.} For $\Xi_i \in \Res_{1}$, one has ${\widehat f}_{n+1}^{R1}= f_{n}f_1^{\circ,R1}$ with $f_1^{\circ,R1}$ being the single particle (equilibrium) density in \fh{reservoir 1} and equivalently ${\widehat f}_{n+1}^{R2}= f_{n}f_1^{\circ,R2}$ for the other reservoir. Or, equivalently, ${\widehat f}_{n+1}^{R1}=f_{R_{1}}^{GC}$ (same for $R_{2}$,) if one makes the modeling choice of the grand canonical distribution for each reservoir.

  Moreover, \fh{the decomposition of the boundary} $\partial \Omega = \partial\Omega^{1} \cup \partial\Omega^{2}$ implies \fh{the splitting of the surface integral}:
\begin{equation}
- (n+1)
        \int\limits_{\partial \Omega} 
          \int\limits_{\reals^3}
            \left(\vec{v}_i \cdot \vn\right) \ {\widehat f}_{n+1}(t, \VX^n, (\vq_i,\vp_i)\,
              d^3 p_i\, d\sigma_i
         = \sum_{r=1}^{2} I_{\partial\Omega^{r}}
\end{equation}
where
\begin{equation}\label{eq:IOmega}
I_{\partial\Omega^{r}}=
        -(n+1) \int\limits_{\partial\Omega^{r}}
          \int\limits_{\reals^3}
            \left(\vec{v}_i \cdot \vn\right) \ {\widehat f}_{n+1}(t, \VX^n, (\vq_i,\vp_i)\,
              d^3 p_i\, d\sigma_i\,.
\end{equation}
In the case $\vec{v}_i \cdot \vn < 0$, one has to replace ${\widehat f}^{n+1}$ with $f_{n}f_1^{\circ,R1}$ in the integral over $\partial\Omega^{1}$ and with $f_{n}f_1^{\circ,R2}$ in the integral over $\partial\Omega^{2}$.
 
\end{enumerate}


After collecting the results of the previous steps, we straightforwardly obtain:
\begin{equation}\label{eq:LiouvilleTypeEquationI}
\frac{\partial f_n}{\partial t} 
+ \fh{\{f_n, H_n\}}
= \sum_{r\in\{R1,R2\}}\left(\Psi_{n,r}+ \Phi_{n,r}^{n+1} \right)
\end{equation}
\fh{where the terms on the r.h.s.\ closely resemble those for a single reservoir and read, e.g., for $r=R1$:}
\begin{equation}\label{eq:LiouvilleHierarchy}
\Psi_{n,R1}[\VX^n, f_n] =
    - \sum\limits_{i=1}^{n} \nabla_{\vec{p}_i}\cdot\Bigl(\Fmean^{R1}(\vq_i) f_n(t, \VX^{i-1}, X_{i}, \VX_{i}^{n-i})\Bigr)
\end{equation}
and
\begin{multline}
  \Phi_{n,R1}^{n+1}[\VX^n, f_n, f_{n+1}] = \\
  (n+1) \int\limits_{\partial \Omega^{1}}
          \int\limits_{(\vec{p}_i \cdot \vn) > 0}
            \hspace{-5pt}
            \left(\vec{v}_i \cdot \vn\right) \ 
            \Bigl( {f}_{n+1}\left(t, \VX^{n},  (\vq_i,\vp_i)\right)
                 - {f}_{n}\left(t, \VX^n\right){f}_{1}^{\circ,R1}\left(\vq_i,-\vp_i\right)
            \Bigr)\,
          d^3 p_i\ d\sigma_i \,.
\end{multline}
The setup of \cref{outofeq} and the marginalization procedure can be straightforwardly extended to an arbitrary number of $m$ disjoint reservoirs, interfaced with $\Omega$:
\begin{equation}
\frac{\partial f_n}{\partial t} 
 + \fh{\{f_n, H_n\}}
= \sum_{r=1}^{m} \left(\Psi_{n,r}+ \Phi_{n,r}^{n+1} \right) ,
\end{equation}  
\fh{which describes a linear and additive action of independent reservoirs.}


\section{Technical details of the simulations}
\label{sec:technicalsection}

The setup of both the AdResS and the full atomistic reference simulations was the same as described in detail in Ref.~\citenum{jcpnoneq} and its supplementary material.
The investigated LJ fluids consist of point particles of mass $m$ that interact via the shifted and smoothly truncated pair potential
$
 U(r) = [U_\text{LJ}(r) - U_\text{LJ}(r_c)] f((r - r_c)/h)
$
for $r \leq r_c$, and $U(r) = 0$ otherwise, with
$
 U_\text{LJ} = 4 \epsilon \bigl[ (r/\sigma)^{-12} - r/\sigma)^{-6} \bigr] ,
$
the cutoff radius $r_c = 2.5\sigma$, the smoothing function $f(x)=x^4/(1+x^4)$, and $h=0.005\sigma$.
The parameters $\epsilon$ and $\sigma$ are taken to define the units for energy and length,
$\tau=\sqrt{m \sigma^2/\epsilon}$ is the unit of time.
Dimensionless quantities are defined as $\rho^* = \rho\sigma^3$ and $T^* = \kB T/\epsilon$ for density and temperature, respectively.
For particle pairs involving at least one tracer particle of AdResS, the interaction is switched off, $\epsilon=0$.

For the simulations reported in \cref{highgrad}, we used reservoir states in the liquid phase along the same isobar, i.e., they have the same pressure $P(T, \rho) = \text{const}$.
The hot reservoir serves as reference state point and is chosen at temperature $T_\text{hot}^* = 0.95$ and density $\rho_\text{hot}^* = 0.622$, 
which results in a (reduced) pressure of $P^* := P \sigma^3/\epsilon \approx 0.045$.
The state points of the cold reservoir were determined such that they are at the same pressure as the hot reservoir. We used the following points in the temperature--density plane:
$(T_2^*, \rho_2^*) = (0.825, 0.72)$,
$(T_3^*, \rho_3^*) = (0.7, 0.791)$,
and $(T_4^*, \rho_4^*) = (0.625, 0.828)$,
leading to temperature differences between the reservoirs of $\Delta T^* = 0.125$, $0.250$, and \roya{$0.325$}, respectively.

For the isobaric results shown in \cref{noneqisoc}, state points along a slightly different isobar were used, namely $(T_\text{hot'}^*, \rho_\text{hot'}^*) = (0.975, 0.5987)$ and
$(T_\text{cold}^*, \rho_\text{cold}^*) = (0.85, 0.7047)$, both at a pressure of $P^*= 0.051$.
The data for the isochoric setup were obtained with reservoir states that represent the same, average density, $\bar\rho^*= (\rho_\text{hot'}^* + \rho_\text{cold}^*)/2 = 0.6517$, but different temperatures
\roya{$T_\text{hot'}^*$} and $T_\text{cold}^*$ as before;
the corresponding pressures differ widely: \fh{$p_\text{hot'}^*=0.194$ and $p_\text{cold}^*=-0.140$.} The negative pressure implies that in equilibrium such a reservoir would phase separate. This is not necessarily the case in the non-equilibrium situation. In our case it is found that the pressure gradient is balanced by a density gradient so that effectively the reservoir is no longer in the unstable state (but at a higher density, i.e., liquid again). So in essence, the example makes sense in the non-equilibrium case and it represents a challenging condition for testing our model.
Both AdResS and reference simulations were performed with the GPU-accelerated simulation software ``HAL's MD package'' \cite{colberg2011,*HALMD}.
For all nonequilibrium simulations, a cuboid domain of size $120\sigma \times 20\sigma \times 20\sigma$ was used for the ``Universe'', with the long edge corresponding to the direction along which molecules change their resolution in AdResS.
Periodic boundary conditions were applied at all faces of the cuboid, and a mirrored setup with in total two AT boxes, four $\Delta$ regions, and two TR regions was employed as in Ref.~\citenum{jcpnoneq}.
The Hamiltonian dynamics of the systems was integrated with the velocity Verlet scheme for a timestep of $0.002\tau$.
No further measures were applied to the AT regions;
the $\Delta$ and TR regions in AdResS and the reservoir regions in the full atomistic reference simulation were thermalized with the Andersen thermostat \cite{andersen}
with the update rate set to $\nu_\text{coupl} = 50\tau^{-1}$ for \cref{highgrad} and $20\tau^{-1}$ for \cref{noneqisoc}.
The AdResS setups contained typically \fh{$16\,000$ LJ particles} on average, whereas about \fh{$31\,000$ LJ} particles were used for the reference simulations.
At each state point, nonequilibrium trajectories over a duration of $15\,000\tau$ each were generated, the first quarter of which ($3\,750\tau$) was discarded for the calculation of stationary time averages.
\fh{For the data analysis, the simulation box was divided at the mirror plane of the setup and the results were averaged over both halves.}
The averages are done over one long trajectory for the isobaric case and over three different trajectories for the isocoric case.


\bibliography{liouvillenoneq}
\end{document}